\documentclass{PoS}

\title{Simulations of supersymmetric Yang-Mills theory}

\ShortTitle{Simulations of SUSY}

\author{K.\ Demmouche$^{a}$,
        F.\ Farchioni$^{a}$,
        A.\ Ferling$^{a}$,
        I.\ Montvay$^{b}$,
        \speaker{G.\ M\"unster}$^{a}$,
        E.E.\ Scholz$^{c}$,
        J.\ Wuilloud$^{a}$\\
\llap{$^{a}$}Universit\"at M\"unster,
  Wilhelm-Klemm-Strasse 9, D-48149 M\"unster, Germany\\
\llap{$^{b}$}Deutsches Elektronen-Synchrotron DESY,
  Notkestr.\,85, D-22603 Hamburg, Germany\\
\llap{$^{c}$}Fermi National Accelerator Laboratory,
  Batavia, IL 60510, USA\\
E-mail: \email{munsteg@uni-muenster.de}}

\abstract{Results of a numerical simulation concerning the low-lying
spectrum of four-dimensional $\mathcal{N}=1$ SU(2) Supersymmetric
Yang-Mills (SYM) theory on the lattice with light dynamical gluinos are
reported. We use the tree-level Symanzik improved gauge action and Wilson
fermions with stout smearing of the gauge links in the Wilson-Dirac
operator. The configurations are produced with the Two-Step Polynomial
Hybrid Monte Carlo (TS-PHMC) algorithm. We performed simulations on lattices
up to a size of $24^3 \cdot 48$ at $\beta=1.6$. Using QCD units with the
Sommer scale being set to $r_0=0.5\,{\rm fm}$, the lattice spacing is about
$a\simeq 0.09\,{\rm fm}$, and the spatial extent of the lattice corresponds
to $2.1$ fm to control finite size effects. At the lightest simulated gluino
mass our results indicate a mass for the lightest gluino-glue bound state,
which is considerably heavier than the values obtained for its possible
superpartners. Whether supermultiplets are formed remains to be studied in
upcoming simulations.}

\FullConference{The XXVII International Symposium on Lattice Field Theory\\
		 July 26-31, 2009\\
		 Peking University, Beijing, China}

\newcommand{\U}{\mathrm{U}}
\newcommand{\SU}{\mathrm{SU}}
\newcommand{\Tr}{\mbox{Tr}}
\newcommand{\I}{\ensuremath{\mathrm{i}}}

\begin{document}

\section{Introduction}
\label{sec:intro}

Supersymmetric theories have received increasing interest in elementary
particle physics. The supersymmetric extension of the Standard Model with
$\mathcal{N}=1$ supercharge and other models beyond the Standard Model have
supersymmetry (SUSY) as an essential ingredient.

The minimal supersymmetric extension of the $\SU (N_{c})$ gauge theory
describing self-interactions of gauge fields $A_{\mu}^a$, corresponding to
the {\em gluons} ($g$), is given by the $\mathcal{N}=1$ Supersymmetric
Yang-Mills (SYM) theory. The supersymmetric partners of the gluons are
described by spin-1/2 Majorana fermion fields $\lambda_{a}\; (a=1\dots
N_{c}^2-1)$, the {\em gluinos} ($\tilde g$). Compatibility of SUSY with
gauge invariance requires that the gluinos transform in the adjoint
representation of the gauge group. This theory describes the interactions
between gluons and gluinos. The Lagrangian of SYM theory in the continuum,
including a SUSY breaking mass term, reads
\begin{equation}
\label{eq: SYM Lag cont}
\mathcal{L}_{SYM}
= -\frac{1}{4} F_{\mu\nu}^a F^{a\mu\nu} 
+ \frac{\I}{2} {\bar\lambda}^a(x) \gamma^\mu (\mathcal{D}_{\mu}\lambda(x))^a
- \frac{m_{\tilde g}}{2}{\bar\lambda}^a\lambda^a \,.
\end{equation}
The mass term here introduces a {\em soft breaking} of supersymmetry.

In the low-energy regime, where the interactions become strong, arguments
based on the low-energy effective Lagrangian approach \cite{VeYa,FaGaSch}
predict the occurrence of non-perturbative dynamics like confinement and
spontaneous chiral symmetry breaking in SUSY gauge theories. Confinement is
realised by colourless bound states. In the case where the last term in
Eq.~(\ref{eq: SYM Lag cont}) is switched off ($m_{\tilde g}=0$), the
anomalous global chiral symmetry $\U (1)_{\lambda}$ is present. The anomaly
does not break the global chiral symmetry completely and a discrete subgroup
$Z_{2N_{c}}$ remains. As in the case of QCD, the discrete chiral symmetry is
expected to be spontaneously broken to $Z_{2}$ by the non-vanishing value of
the gluino condensate $\langle {\bar\lambda} \lambda\rangle$. The
consequence of this spontaneous breaking is the existence of $N_{c}$
degenerate ground states with different orientations of the gluino
condensate.

Another interesting aspect of SYM is its equivalence to QCD with a single
quark flavour ($N_{f}=1$ QCD) in the limit of a large number of colours
($N_{c}\rightarrow \infty$), where the Majorana spinor is replaced by the
single Dirac spinor \cite{{ArmoniShifmanVeneziano}}. The latter model is
also object of investigation by our collaboration \cite{nf1}.

A lattice formulation of SYM suitable for numerical simulations, employing
Wilson fermions, has been proposed by Curci and Veneziano \cite{CuVe}. First
non-perturbative investigations of SYM on the lattice using this formulation
have been performed by the DESY-M\"unster-Roma collaboration; for a review
see Ref.~\cite{Montvay_SYM} and references
\cite{CaetAl,DoetAl,fed-peetz-res,Demmouche:2008aq,Demmouche:2008ms}. SUSY
is broken explicitly by the lattice discretisation. In the Wilson approach
the mass term and the Wilson-term break both chirality and SUSY. Both
symmetries are expected to be recovered in the continuum limit by tuning the
relevant bare mass term to its critical value corresponding to a massless
gluino ($m_{\tilde g}=0$), and the gauge coupling towards zero.

In recent years, simulations of $\mathcal{N}=1$ SYM on the lattice using
Ginsparg-Wilson fermions with good chiral properties, such as domain wall
fermions, have been initiated
\cite{Endres:2009yp,Giedt:2008xm,Fleming:2000fa}. For large lattice volumes
and small lattice spacings these formulations require, however, a
significantly larger amount of computing resources than the Wilson
formulation. The gain of no need for tuning the position of the zero gluino
mass point does not compensate by far the advantage of Wilson fermions.

Here we report on recent work continuing the project of the DESY-M\"unster
collaboration for the simulation of $\mathcal{N}=1$ $\SU (2)$ SYM. We
present results in the light of the newly used TS-PHMC algorithm and
improved actions.

The most important characteristics of the theory is the mass spectrum of
bound states, for which the low-energy effective theories predict a
reorganisation of the masses in two massive Wess-Zumino supermultiplets at
the SUSY point \cite{VeYa,FaGaSch}, where the soft breaking vanishes. The
introduction of a small gluino mass removes the mass degeneracy between the
supermultiplet members.

%%%%%%%%%%%%%%%%%%%%%%%%%%%%%%%%%%%%%%%%%%%%%%%%%%%%%%%%%%%%%%%%%%%%%%%%
\section{Lattice formulation of $\mathcal{N}=1$ SYM theory}
\label{sec:lattice formulation}

For the gauge fields we employ the {\em tree-level improved Symanzik}
(tlSym) action. The gluinos are represented by Majorana fermions $\lambda^a$
in the adjoint representation.  The fermion part of the Curci-Veneziano
action, describing the gluinos, is given by
\begin{equation}
S^W_{\tilde g} 
= \frac{1}{2} \sum_x \bar{\lambda}(x)\lambda(x) 
- \frac{\kappa}{2} \sum_x \sum_\mu
[\bar{\lambda}(x + \hat{\mu}) V_\mu(x) (1 + \gamma_\mu) \lambda(x) 
%\nonumber \\ 
+ \bar{\lambda}(x) V_\mu^T(x) (1 - \gamma_\mu) \lambda(x+\hat{\mu})] \,,
\label{eq:defLatticeFermionAction} 
\end{equation}
where $\kappa$ is the bare hopping parameter which encodes the bare gluino
mass $\kappa=(2m_{{\tilde g},0}+8)^{-1}$.  The real orthogonal matrices
$V_{\mu}(x)$ are the gauge links in the adjoint representation.

The links $U_{x,\mu}$ in the fermion action can be replaced by {\em
stout}-smeared links \cite{MorningstarPeardon}. This has the advantage that
short range {\em topological defects} of the gauge field and the
corresponding small eigenvalues of the fermion matrix are removed. We prefer
to keep the action well localised and hence only perform a single
stout-smearing step.

Similarly to QCD, the mass term proportional to $m_{{\tilde g},0}$ breaks
chirality. In the present case it also breaks the supersymmetry. A massless
gluino, $m_{\tilde g}=0$, is obtained by tuning the relevant bare mass term
to its critical value ($m_{0}\rightarrow m_{0cr})$ or equivalently
$\kappa\rightarrow\kappa_{cr}$.

The fermion action can be rewritten in term of an antisymmetric matrix $M=
\mathcal{C} Q$, where $Q$ is the non-hermitian fermion matrix or lattice
Wilson-Dirac operator for Dirac fermions in the adjoint representation, 
and $\mathcal{C}$ is the charge conjugation matrix in the spinorial 
representation. Integration of the fermionic variables yields the 
Pfaffian of $M$, whose absolute value equals the square root of the
fermion determinant. Effectively, this corresponds to a flavour number
$N_{f}=\frac{1}{2}$. In the Wilson setup the Pfaffian can become negative
even for positive gluino masses.

In our numerical simulations we include the dynamics of the gluino by the
Two-Step Polynomial Hybrid Monte Carlo (TS-PHMC) \cite{MontvPHMC} algorithm
with flavour number $N_{f}=\frac{1}{2}$. This has the consequence that only
the absolute value of the Pfaffian is taken into account in the updating of
the gauge field configuration. The sign of the Pfaffian has to be included
in a reweighting step when calculating expectation values.
It can be shown that the sign of the Pfaffian is equal to the sign of the 
product of half of the doubly degenerate negative real eigenvalues of $Q$.
For positive gluino masses sufficiently far away from zero a negative sign
of the Pfaffian rarely occurs in the updating sequence and therefore a sign
problem does not show up.

The TS-PHMC algorithm turned out to be very efficient in producing short
autocorrelations among the gauge configurations. For instance, in the
stout-smeared runs on a $24^3\cdot 48$ lattice the integrated
autocorrelation of the average plaquette (which belongs to the worst
quantities from the point of view of autocorrelations) did always satisfy
$\tau_{int}^{plaq} < 10$.

The values of the gauge coupling parameter $\beta$ can be fixed by
investigating the static potential of an external fundamental charge and
extracting the Sommer scale parameter $r_0/a$ \cite{Sommer}. Note that in
analogy with QCD, we set the value of $r_0$ by definition to $r_0=0.5\,{\rm
fm}$. In this way we can use familiar QCD units for physical dimensionful
quantities.

The TS-PHMC runs were done on $16^3 \cdot 32$ and $24^3 \cdot 48$ lattices
at $\beta = 1.6$ at various values of the hopping parameter $\kappa$. On the
$24^3 \cdot 48$ lattice four points have been simulated with stout-links.
The lattice spacing amounts to $a\simeq 0.09\,{\rm fm}$. The lattice
extension $L\simeq 2.1\,{\rm fm}$ is expected to be large enough to allow
control over finite volume effects on the bound state masses.

The dimensionless quantity $M_{r}\equiv (r_{0}m_{\pi})^2$, where $m_{\pi}$
is the pion mass in the corresponding theory with two Dirac fermions in the 
adjoint representation, is expected to be proportional to the gluino mass
(see below), and can be considered to be a measure of it.

An issue in lattice simulation is the lightness of the dynamical quarks
which leads to slowing down of the update algorithms. With the new TS-PHMC
algorithm the lightest adjoint pion mass in our simulations was about
440 MeV. Simulations for smaller gluino masses and/or finer lattice 
spacings are going on presently.

%%%%%%%%%%%%%%%%%%%%%%%%%%%%%%%%%%%%%%%%%%%%%%%%%%%%%%%%%%%%%%%%%%%%%%%%
\section{Low-lying bound state spectrum}
\label{sec:Low-lying masses determination}

For the investigation of the low-lying bound state spectrum we concentrate
on the projecting operators employed for the construction of the low-energy
Lagrangians of~\cite{VeYa} and \cite{FaGaSch}. These are expected to
dominate the dynamics of SYM at low energies. Previous experience on the
field can be found in~\cite{CaetAl} and \cite{fed-peetz-res}. We investigate
spin-0 gluino-gluino bilinear operators (adjoint mesons), a spin-1/2 mixed
gluino-glue operator and spin-0 glueball operators. In some cases smearing
techniques such as APE~\cite{Alb87} and Jacobi smearing~\cite{Jacobi} are
applied in order to increase the overlap of the lattice operator with the
low-lying bound state.

The adjoint mesons are colourless composite states of two gluinos with
spin-parity $0^+$ and $0^-$. In analogy to flavour singlet states in QCD
we denote the former $a$-$\eta^\prime$ and the latter $a$-$f_0$, where the
prefix $a$ stays for ``adjoint''. The associated projecting operators are 
the gluino bilinear operators ${\cal O}_{mes}=\bar\lambda\Gamma\lambda$
where $\Gamma = \gamma_5,1$ respectively. 

The resulting propagator consists of connected and disconnected
contributions. The exponential decay of the connected term defines the
adjoint pion mass $m_{a\mbox{-}\pi}$. This quantity, even if not associated
to a physical state of SYM, can be used to characterise the gluino mass.
Indeed, according to arguments involving the OZI-approximation of SYM, the
adjoint pion mass is expected to vanish for a massless gluino and the
behaviour $m^2_{a\mbox{-}\pi}\propto m_{\tilde g}$ can be assumed for light
gluinos~\cite{VeYa,fed-peetz-res}.

For the positive parity glueball $0^+$ we adopted the simplest interpolating
operator built from space-like plaquettes. In order to improve the signal we
applied APE smearing with the variational method~\cite{Variational}.

Also here, as for the scalar $a$-$f_{0}$, the gauge samples generated with
stout links turn out to generally give better results for the glueball
masses. 

The gluino-glueballs ($\tilde g\mbox{-}g$) are spin-$\frac{1}{2}$ colour
singlet states of a gluon and a gluino. They are supposed to complete the
Wess-Zumino supermultiplet of the adjoint mesons \cite{VeYa}. We adopt for
this state the lattice version of the gluino-glue operator
$\Tr_c[F\sigma\lambda]$~\cite{VeYa}, where the field-strength tensor
$F_{\mu\nu}(x)$ is replaced by the clover-plaquette operator
$P_{\mu\nu}(x)$~\cite{FaetAl,fed-peetz-res}. We apply APE smearing for the
links and Jacobi smearing for the fermion fields in order to optimise the
signal-to-noise ratio and to obtain an earlier plateau in the effective
mass.

A vanishing gluino mass is a prerequisite for supersymmetry in the continuum
limit. Therefore the point corresponding to a massless gluino is of high
interest in lattice simulations of SYM. With Wilson fermions this point must
be located by tuning the hopping parameter. The subtracted gluino mass can
be determined in a direct way from the study of lattice SUSY Ward-Identities
(WIs)~\cite{FaetAl} and, in an indirect way, from the vanishing of the
adjoint pion mass. Indeed, as mentioned above, the pion mass squared
$(am_{\pi})^2$ is expected to vanish linearly with the (renormalised) gluino
mass. Both the WIs and adjoint pion mass methods give consistent estimates
of the critical hopping parameter $\kappa_{cr}$ corresponding to vanishing
gluino mass.

\begin{figure}[!htb]
\centering
\includegraphics[width=0.56\textwidth]{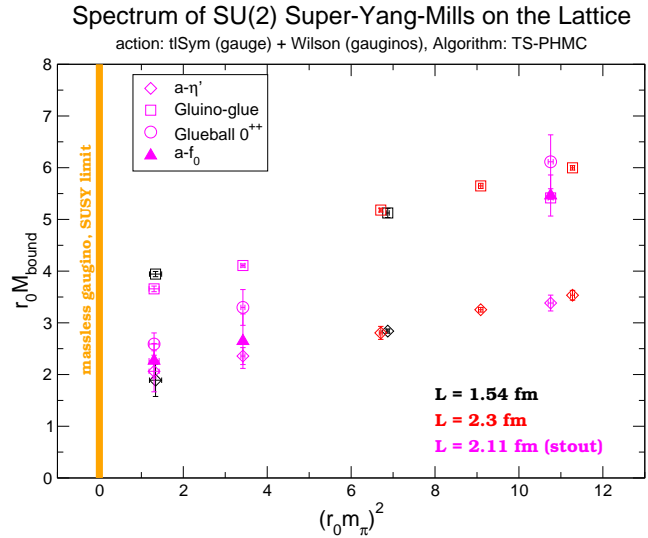}
\parbox[t]{0.8\textwidth}{%
\caption{\label{spectrum}%
Low-lying bound state masses of ${\cal N}=1$ SU(2) SYM as a function of the 
adjoint pion mass squared (physical units).}
}
\end{figure}

The masses of the lightest bound states of low-energy $\mathcal{N}=1$ SYM
determined in this work are shown in Fig.~\ref{spectrum}. The masses of
$a$-$f_0$ and the glueball from the unstout ensembles have a very low
signal-to-noise ratio and are not shown. The results at the smallest gluino
mass are preliminary.

The masses in Fig.~\ref{spectrum} are multiplied by the Sommer scale
parameter and plotted as a function of the squared adjoint pion mass. The
lightest simulated adjoint pion mass is about 440 MeV in our units.  The
vertical line in Fig.~\ref{spectrum} highlights the massless gluino limit
where SUSY restoration is expected up to $\mathcal{O}(a)$ effects.

The bound state masses appear to be characterised by a linear dependence on
$(r_{0}m_{\pi})^2$. The gluino-glueball $({\tilde g}g)$ with a mass of about
1440 MeV (in our unit where $r_0=0.5$fm) turns out to be considerably heavier
than the $a$-$\eta^\prime$ with a mass of 810 MeV. Furthermore, the masses
of the scalar glueball and the scalar meson $a$-$f_{0}$ converge to a common
point with the pseudoscalar near the region where SUSY is expected. The
behaviour of scalars is compatible with mixing between $0^+$ glueball and
$a$-$f_{0}$. The behaviour of masses suggests a lower supermultiplet, while
the spin-1/2 candidate remains heavier up to the smallest simulated gluino
mass in this simulation. Whether this outcome is a discretisation artefact
or a physical effect, as claimed in \cite{BeMi}, should become clear in 
future studies at finer lattice spacings. As the data at small gluino mass 
are preliminary, it would be premature to draw conclusions.

%%%%%%%%%%%%%%%%%%%%%%%%%%%%%%%%%%%%%%%%%%%%%%%%%%%%%%%%%%%%%%%%%%%%%%%%
\section{Spontaneous symmetry breaking}

An important feature of the strong interaction dynamics of SYM theory is the
spontaneous breaking of the discrete chiral symmetry. The appearance of two
degenerate ground states in case of SU(2) gauge group was observed in
\cite{discrete} on a small ($6^3 \cdot 12$) lattice by neglecting the sign
of the Pfaffian.

\begin{figure}[!htb]
\centering
\includegraphics[width=0.5\textwidth,angle=270]{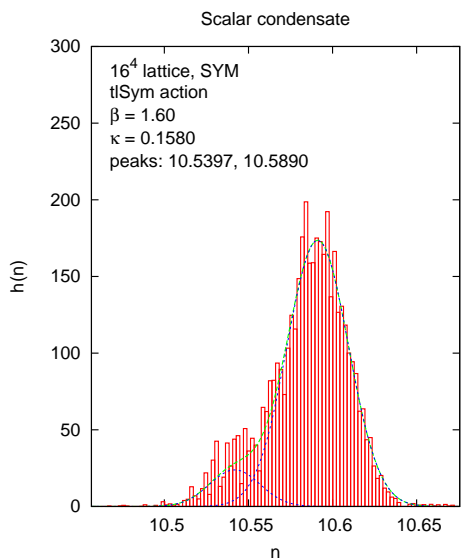}
\parbox[t]{0.8\textwidth}{%
\caption{\label{discrete}%
Distribution of the gluino condensate at $\beta=1.6,\;\kappa=0.158$.}
}
\end{figure}

Recently we repeated this computation on a larger ($16^4$) lattice and also
taking into account the Pfaffian signs. The observed distribution of the
gluino condensate is shown in Fig.~\ref{discrete}.

The distribution cannot be fitted with a single Gaussian, but a two-Gaussian
fit shown on the figure, corresponding to the two ground states, is
reasonable. From the difference of the positions of the two peaks one can,
in principle, determine the magnitude of the gluino condensate. This
however requires the knowledge of renormalisation constants and therefore is
left for a future work.

%%%%%%%%%%%%%%%%%%%%%%%%%%%%%%%%%%%%%%%%%%%%%%%%%%%%%%%%%%%%%%%%%%%%%%%%
\section{Conclusion}
\label{sec:conclusion}

In this work first {\em quantitative} results of the low-energy spectrum of
${\cal N}=1$ supersymmetric Yang-Mills theory are obtained. Physical volumes
larger than 2 fm have been simulated, which is the volume usually required 
for spectrum studies in lattice gauge theory. The comparison of masses on 
different volumes in otherwise same conditions reveals negligible finite 
size effects at least for moderate gluino masses. We have collected
higher statistics and have used efficient dynamical algorithms such as
TS-PHMC, which is suitable for light fermion masses. In addition, the
supersymmetric Ward identities and other observables like the confinement
potential and the gluino condensate have been investigated.

From the results of the mass spectrum the question of the gluino-gluino and
gluino-glueball mass splitting remains open. It can only be answered by
further simulations allowing an extrapolation to the continuum limit.


\begin{thebibliography}{99}
%
\bibitem{VeYa}
G.~Veneziano and S.~Yankielowicz,
Phys.\ Lett.\ B {\bf 113} (1982) 231.
%
\bibitem{FaGaSch}
G.R.~Farrar, G.~Gabadadze and M.~Schwetz,
Phys.\ Rev.\ D {\bf 58} (1998) 015009.
%
\bibitem{ArmoniShifmanVeneziano}
A.~Armoni, M.~Shifman and G.~Veneziano,
 %`From Super-Yang-Mills theory to QCD: planar equivalence and
 % its implications,''
in \textit{From Fields to Strings: Circumnavigating Theoretical Physics},
vol.~1, eds.\ M.~Shifman, A.~Vainshtein, J.~Wheater,
World Scientific, Singapore, 2005, p.~353; [hep-th/0403071].
%
\bibitem{nf1}
F.~Farchioni, G.~M\"unster, T.~Sudmann, J.~Wuilloud, I.~Montvay 
and E.~E.~Scholz,
PoS(LATTICE 2008) 128, 
PoS(LATTICE 2007) 135, 
Eur.\ Phys.\ J.\ C {\bf 52} (2007) 305.
%
\bibitem{CuVe}
G.~Curci and G.~Veneziano,
Nucl.\ Phys.\ B {\bf 292} (1987) 555.
%
\bibitem{Montvay_SYM}
I.~Montvay,
Int.\ J.\ Mod.\ Phys.\ A {\bf 17} (2002) 2377.
%
\bibitem{CaetAl}
I.~Campos, A.~Feo, R.~Kirchner, S.~Luckmann, I.~Montvay, G.~M\"unster,
K.~Spanderen and J.~Westphalen,
Eur.\ Phys.\ J.\ C {\bf 11} (1999) 507.
%
\bibitem{DoetAl}
A.~Donini, M.~Guagnelli, P.~Hernandez and A.~Vladikas,
Nucl.\ Phys.\ B {\bf 523} (1998) 529.
%
\bibitem{fed-peetz-res}
F.~Farchioni and R.~Peetz, 
Eur.\ Phys.\ J.\ C \textbf{39} (2005) 87.
%
\bibitem{Demmouche:2008aq}
K.~Demmouche, F.~Farchioni, A.~Ferling, G.~M\"unster, J.~Wuilloud, 
I.~Montvay and E.~E.~Scholz, 
PoS(Confinement 2008) 136.
%  
\bibitem{Demmouche:2008ms}
K.~Demmouche, F.~Farchioni, A.~Ferling, G.~M\"unster, J.~Wuilloud, 
I.~Montvay and E.~E.~Scholz, 
PoS(LATTICE 2008) 061.
%
\bibitem{Endres:2009yp}
M.G.~Endres,
PoS(LATTICE 2008) 025.
%``Dynamical simulation of N=1 supersymmetric Yang-Mills theory with domain
%wall fermions,''
%
\bibitem{Giedt:2008xm}
J.~Giedt, R.~Brower, S.~Catterall, G.~T.~Fleming and P.~Vranas,
%``Lattice super-Yang-Mills using domain wall fermions in the chiral limit,''
Phys.\ Rev.\  D {\bf 79} (2009) 025015.
%
\bibitem{Fleming:2000fa}
G.T.~Fleming, J.~B.~Kogut and P.~M.~Vranas,
%``Super Yang-Mills on the lattice with domain wall fermions,''
Phys.\ Rev.\  D {\bf 64} (2001) 034510.
%
\bibitem{MorningstarPeardon}
C.~Morningstar and M.J.~Peardon, 
Phys.\ Rev.\ D \textbf{69} (2004) 054501.
%
\bibitem{MontvPHMC}
I.~Montvay and E.E.~Scholz, 
Phys.\ Lett.\ B \textbf{623} (2005) 73.
%
\bibitem{Sommer}
R.~Sommer,
Nucl.\ Phys.\  B {\bf 411} (1994) 839.
%
\bibitem{Alb87}
M.~Albanese et al., 
Phys.\ Lett.\ B {\bf 192} (1987) 163.
%
\bibitem{Jacobi}
C.~R.~Allton et al. [UKQCD Collaboration],
Phys.\ Rev.\ D {\bf 47} (1993) 5128.
%
\bibitem{Variational}
M.~L\"uscher and U.~Wolff,
Nucl.\ Phys.\ B {\bf 339} (1990) 222.
%
\bibitem{FaetAl}
F.~Farchioni, A.~Feo, T.~Galla, C.~Gebert, R.~Kirchner, I.~Montvay,
G.~M\"unster and A.~Vladikas,
Eur.\ Phys.\ J.\ C {\bf 23} (2002) 719.
%
\bibitem{BeMi}
L.~Bergamin and P.~Minkowski, hep-th/0301155.
%
\bibitem{discrete}
R.~Kirchner, S.~Luckmann, I.~Montvay, K.~Spanderen and J.~Westphalen,
Phys.\ Lett.\ B {\bf 446} (1999) 209.
%
\end{thebibliography}
\end{document}